\documentclass[english,prc,preprint,showpacs,amsmath,amssymb, floatfix]{revtex4}
\usepackage[T1]{fontenc}
\usepackage[latin1]{inputenc}

\makeatletter

\usepackage{graphicx}
\usepackage{array}
\usepackage{longtable}
\usepackage{babel}
\usepackage{subfigure}

\def\beq{\begin{equation}}
\def\eeq{\end{equation}}
\def\ba{\begin{eqnarray}}
\def\ea{\end{eqnarray}}

\def\v8p{v_8^\prime}

\newcommand{\boldnabla}{\mbox{\boldmath$\nabla$}}

\newcommand{\ran}{\rangle}
\newcommand{\lan}{\langle}
\newcommand{\bec}{\begin{center}}
\newcommand{\enc}{\end{center}}
\newcommand{\bit}{\begin{itemize}}
\newcommand{\eit}{\end{itemize}}

\newcommand{\rta}{\rightarrow}

\newcommand{\bfk}{{\bf k}}

\newcommand{\bfr}{{\bf r}}

\newcommand{\calF}{{\cal F}}
\newcommand{\calG}{{\cal G}}

\newcommand{\calM}{{\cal M}}
\newcommand{\calN}{{\cal N}}

\newcommand{\calS}{{\cal S}}

\newcommand{\barn}{\overline{n}}

\usepackage{babel}
\makeatother

\begin{document}

\title{Variational Theory of Hot Nucleon Matter}

\author{Abhishek Mukherjee\cite{am} and V. R. Pandharipande\cite{vrp} }

\affiliation{Department of Physics, University of Illinois at Urbana-Champaign,
\\
1110 W. Green St., Urbana, IL 61801, U.S.A.}

\date{\today{}}

\begin{abstract}
We develop a variational theory of hot nuclear matter in neutron stars
and supernovae. It can also be used to study charged, hot nuclear
matter which may be produced in heavy-ion collisions. This theory
is a generalization of the variational theory of cold nuclear and neutron
star matter based on realistic models of nuclear forces
and pair correlation operators. The present approach uses microcanonical
ensembles and the variational principle obeyed by the free energy.
In this paper we show that the correlated states of the microcanonical
ensemble at a given temperature $T$ and density $\rho$ can be orthonormalized
preserving their diagonal matrix elements of the Hamiltonian. This allows for
the minimization of the free energy without corrections from the nonorthogonality
of the correlated basis states, similar to that of the ground state energy.
Samples of the microcanonical ensemble
can be used to study the response, and the neutrino luminosities
and opacities of hot matter. We present methods to orthonormalize the correlated
states that contribute to the response of hot matter.
\end{abstract}


\maketitle

\section{Introduction}

\emph{Ab initio} theories of strongly interacting hot matter are extremely
challenging. In principle the properties of hot matter can be calculated
starting from a realistic Hamiltonian with the path integral Monte
Carlo method \cite{pimcb}. Calculations are practical for simple
systems of interacting spin zero bosons such as atomic $^{4}$He liquids
and solids \cite{pimche41, pimche42,pimche43}. They become more difficult even for simple
systems of fermions interacting by spin independent potentials, such
as atomic liquid $^{3}$He \cite{pimche3}, and hydrogen plasma \cite{pimchp}
due to the fermion sign problem. The path integral Monte Carlo treatment
is expected to become much more difficult due to the strong spin-isospin
dependence of nuclear forces and their tensor and spin-orbit components.
In the traditional Monte Carlo approaches these complexities of the
nuclear forces make computations more expensive by a factor $\geq2^{A}$,
where $A$ is the number of nucleons and the equality applies for
pure neutron matter. With the present state of the art computing facilities
traditional quantum Monte Carlo calculations have been carried out
for cold neutron matter using a periodic box containing 14 neutrons
\cite{pnmgfmc}. Attempts are also being made to eliminate this $2^{A}$
factor using the auxiliary field diffusion Monte Carlo method \cite{afdmc},
however the fermion sign problem is more acute for this method, and
applications have been limited to cold pure neutron matter \cite{pnmafdmc}.

Cold nuclear matter has traditionally been studied with variational methods \cite{ap,apr}
and Brueckner theory \cite{brusnm,brupnm}. There is close agreement
between these two methods, and comparison with the essentially exact
Green's function Monte Carlo calculations suggests that the errors
in present variational calculations of pure neutron matter are only
$\sim$ 8 \% at densities $\leq\rho_{0}=0.16$ fm$^{-3}$ \cite{pnmgfmc}.
In the case of symmetric nuclear matter the errors have been estimated
to be $<$ 10 \% \cite{mpr}. 

In this paper we develop the formalism for a variational theory
for nuclear matter at finite temperature using correlated basis states (CBS)
defined in the next subsection.  The correlated basis states and the  thermodynamic variational
principle used to calculate the free energy of matter is discussed in
the following subsections.  In these subsections we review the scheme suggested
 in Ref.~\cite{SP} to develop a variational theory of hot matter, and comment
on the concerns expressed in its early applications \cite{FPS,FP81}
due to the nonorthogonality of the CBS.

In Section II we show that these problems can be resolved if one works
in a microcanonical ensemble.  We show that there are no orthogonality corrections to the free
energy in this scheme.  In Section III we consider the CBS that contribute to the
response of the hot matter, and conclude in Section IV.  

\subsection{Correlated Basis States}

Let the stationary states of a non interacting Fermi gas be denoted by
$|\Phi_{I}\{ n_I(\bfk,\sigma_z)\}\ran$,
where $\{ n_{I}(\bfk,\sigma_z)\}$ are the occupation numbers of single particle states
labeled with momentum $\bfk$ and spin projection $\sigma_z$, in the many-body state $I$.
The single-nucleon states of a non interacting nucleon gas have isospin $\tau_z$ as an
additional quantum number.  We have suppressed it here for brevity.  
For each of the states $I$, we can construct a normalized correlated
basis state (CBS) \cite{Feenberg,Clark,PWRMP} which is conventionally defined
as:
\beq
|\Psi_{I})=\frac{\calG|\Phi_{I}\ran}
{\sqrt{\lan \Phi_{I}|\calG^{\dagger}\calG|\Phi_{I}\ran}},
\eeq
where $\calG$ is a many-body correlation operator.  Many problems in the variational
theory of strongly interacting quantum liquids originate from the fact that useful
forms of $\calG$ are not unitary operators.  In recent studies  $\calG$   
has been approximated by a symmetrized product of pair correlation
operators $\calF_{ij}$ \cite{PWRMP,apr,mpr} where $i$ and $j$ label the nucleons:
\beq
\calG = \calS \prod_{i<j} \calF_{ij}~.
\eeq
Here $\calS$ stands for the symmetrization of the product of the pair correlation operators.
 In the present work we will assume this form of $\calG$, however, improvements such
as the inclusion of three-body correlations can be easily accommodated.
The CBS obtained with this $\calG$ are not orthogonal to each other.
The bras and kets with rounded parenthesis, $(~|$ and $|~)$ are used to denote these
non orthogonal states, while the standard $\lan |$ and $| \ran$ imply orthonormal
states.  

At zero temperature the parameters of $\calG$ or $\calF_{ij}$ are determined
variationally by minimizing the expectation value of the
Hamiltonian, $H$, containing realistic interactions, in the ground
state of the correlated basis $\left|\Psi_0 \right)$ obtained from
the Fermi-gas ground state $| \Phi_0 \ran$.  The CBS are
assumed to provide a good approximation for the stationary states
of the interacting system. Note that this is
in accordance with an important assumption of the Landau theory of
Fermi liquids, i.e. the stationary states (at least the low lying ones)
of an interacting, normal Fermi liquid can be written in one to one
correspondence with those of the non interacting one.

If the $n_0(\bfk,\sigma_z)$ be the occupation numbers of the single particle states in the ground state of the free Fermi gas,
The $\left|n_I(\bfk,\sigma_z)-n_0(\bfk,\sigma_z)\right|$ can be interpreted as
the quasi-particle $(k > k_F)$ and quasi-hole $(k < k_F)$ occupation numbers
of the CBS $\left| \Psi_I \right)$.  
 When the number of quasi-particles
is finite the energy of the state $I$, $E_I$ can be expressed as the
sum of the ground state
energy $E_0$, and a sum of quasi-particle and hole energies:
\beq
\label{eqei}
E_I=\left(\Psi_I|H|\Psi_I\right)=E_0 + \sum_{\bfk,\sigma_z} [n_I(\bfk,\sigma_z)
  -n_0(\bfk,\sigma_z)] \epsilon_0(\bfk,\sigma_z)~.
\eeq
We assume that both the number of particles $N$ and the volume of the liquid $\Omega$,
go to  $\infty$ at a fixed finite density $\rho=N/\Omega$.  The density of quasi-particles
and holes goes to zero when their number is finite.  The single particle energies
$\epsilon_0(\bfk,\sigma_z)$, have significant dependence on $\bfk$ near $k_F$ at low temperature, 
in addition to that absorbed in the $\frac{k^2}{2m*}$ term \cite{FPS},
with $m*$ being the effective mass of the quasiparticle. They are
difficult to calculate {\it ab initio}.

A correlated basis perturbation theory (CBPT) can be developed using
the non orthogonal CBS \cite{Feenberg, Clark} to
study various properties of quantum liquids \cite{FFP, FPRF} at zero temperature.  Much later in the development of
CBPT, a scheme to orthonormalize the CBS preserving their one to one correspondence
with the Fermi gas states and the validity of Eq.~(\ref{eqei}) was found \cite{FP}.  It
simplifies CBPT considerably.

The difference between the internal energies of a liquid at $T > 0$ and $T=0$ is
extensive, {\it i.e.} proportional to $N$, and thus infinite in the thermodynamic limit.  
This implies that at $T>0$ there is an extensive number of quasi-particle excitations,
and the orthonormalization scheme of Ref.~\cite{FP} can not be used without modifications.
The present work can also be considered as an extension of that orthonormalization
scheme to hot matter.  At very small temperatures, the density of quasi-particles is
small, and the $T=0$ formalism can be used neglecting the interaction between quasi-particles,
as in Landau's theory.  However, the domain of the applicability of that approach is very
small \cite{FPS}.

\subsection{The Thermodynamic Variational Principle}

Let $F(T)$ be the free energy of a quantum many body system at
temperature $T$. All other arguments such as the  density $\rho$ and spin-isospin
polarizations etc. have been  suppressed  for brevity. 
The Gibbs-Bogoliubov thermodynamic inequality \cite{Feynman} states that
\beq
\label{thvar}
F(T)  \leq  \textrm{Tr} (\rho_V H)-TS_{V}(T)~, 
\eeq
where $\rho_V$ is any arbitrary density matrix (not to be confused with the density of the system $\rho = N/\Omega$) satisfying 
\beq
\label{trdenmat}
\textrm{Tr}\rho_V = 1
\eeq
 and $S_{V}(T)$ is the
entropy of the density matrix $\rho_V$ at temperature $T$.
The equality holds when $\rho_V$ is the true density matrix of the system. Typically $\rho_V$ is chosen to have the canonical form,
\beq
\label{rhocan}
\rho_{\textrm{can}} = \frac{\exp(-\beta H_V)}{\textrm{Tr}\exp(-\beta H_V)}~,
\eeq
where $\beta$ is the inverse temperature and  $H_{V}$ is chosen as a
suitable, simple and variable variational Hamiltonian.
In this  case Eq.~(\ref{thvar}) becomes
\beq
F(T)  \leq  \frac{\textrm{\textrm{Tr}}(\exp(-\beta H_{V})H)} {\textrm{\textrm{Tr}}(\exp(-\beta H_{V}))}-TS_{V}(T)~,
\eeq

The minimum value of
\beq
\label{freee}
 \frac{\textrm{\textrm{Tr}}(\exp(-\beta H_{V})H)} {\textrm{\textrm{Tr}}(\exp(-\beta H_{V}))}-TS_{V}(T)~,
\eeq
obtained by varying $H_V$, provides an upper bound to the free-energy $F(T)$.

Schmidt and Pandharipande (Ref.~\cite{SP}, henceforth denoted by SP) proposed to use this
variational principle to calculate properties of hot quantum liquids.
They essentially ignored the nonorthogonality of the CBS and assumed
that they are the eigenstates of $H_V$:
\beq
\label{nonoest}
H_V |\Psi_I\{n_I(\bfk,\sigma_z)\} ) = \left[ \sum_{\bfk,\sigma_z} n_I(\bfk,\sigma_z)
   \epsilon_V(\bfk,\sigma_z) \right] |\Psi_I\{n_I(\bfk,\sigma_z)\} )~.
\eeq
The eigenvalues of this $H_V$ can be varied by changing the single-particle energy spectrum
$\epsilon_V(\bfk,\sigma_z)$ and the eigenfunctions by varying the correlation operator $\calG$,
or the pair correlation operators $\calF_{ij}$.  Note that the single particle energies
depend on $\tau_z$, $\rho$, $T$ etc.  but these dependencies are suppressed here.

$H_V$ has the spectrum of a one body Hamiltonian, since its eigenvalues depend only
on the occupation numbers $n_I(k,\sigma_z)$.  It can therefore be easily solved.
At temperature $T$ the average occupation number of a single-particle state is
given by
\beq
\label{barn}
\barn(\bfk,\sigma_z = \pm 1)=\frac{1}{e^{\beta(\epsilon(\bfk,\sigma_z)-\mu_{\pm})}+1}~, 
\eeq
where the chemical potential $\mu_{\pm}$ is required to satisfy
\beq
\rho_{\pm} = \int \frac{d^3k}{(2\pi)^3}~\barn(\bfk,\sigma_z=\pm 1)~.
\eeq

In the above equation $\rho_{\pm}$ is the density of particles with $\sigma_z = \pm 1$.
The entropy $S_V(\rho,T)$ is given by \cite{entropy},
\beq
\label{ent}
S_V(\rho,T)= -k_B \Omega \sum_{\sigma_z} \int \frac{d^3k}{(2\pi)^3}~
          \big[ \barn(\bfk,\sigma_z)~\ln (\barn(\bfk,\sigma_z))  
          +(1- \barn(\bfk,\sigma_z))~\ln(1- \barn(\bfk,\sigma_z))\big]~.
\eeq
where $k_B$ is Boltzmann's constant.

Since the CBS are not mutually orthogonal, Eq.~(\ref{ent}) is only
an approximation if the variational Hamiltonian, $H_V$, is defined by 
Eq.~(\ref{nonoest}). Eq.~(\ref{ent}) will be exact if orthonormalized
correlated basis states (OCBS) are used instead of the non orthogonal
CBS. If all the CBS are orthonormalized by a democratic procedure (like the
L\"owdin method)~ \cite{Lowdin}, which treats all the CBS equally, the diagonal
matrix elements of the Hamiltonian $H$ change by an extensive ($\propto N$)
quantity. 

The diagonal matrix elements of the Hamiltonian, $H$, can be
evaluated using the standard techniques of cluster expansion and chain
summation; these techniques have been developed and studied
extensively in the variational theories of cold (zero temperature)
quantum liquids. On the other hand, if all the CBS are orthonormalized
using the democratic procedure mentioned above, the diagonal matrix
elements of the Hamiltonian, $H$, in the corresponding OCBS are more
difficult to evaluate systematically because of the extensive ($\propto N$)
orthogonality corrections. As such, the variational theory of hot
(finite temperature) quantum liquids so developed using the
orthonormalization scheme discussed above, loses much of the
simplicity of the corresponding zero temperature theory.

At zero temperature a similar problem was addressed by identifying the
ground state and the excitations about the ground state
with a finite number of quasiparticles and quasiholes, as the important
states which contribute to the equilibrium properties and linear
response of cold quantum liquids. It was shown in Ref.~\cite{FP} that
a combination of democratic (L\"owdin) and sequential (Gram-Schmidt)
orthonormalization methods can be used to orthonormalize the CBS, such
that the diagonal matrix elements of the Hamiltonian, $H$, are left
unchanged, in the ground state and in the quasiparticle-quasihole
excitations from the ground state. 

At a finite temperature, the many body states which contribute to the
equilibrium properties (free energy, specific heat etc.) and linear
response of a quantum liquid are the many body states in the
microcanonical ensemble at the corresponding temperature and the
quasiparticle-quasihole excitations from them. 
(Zero temperature is a special case when the
microcanonical ensemble consists of just one state viz. the ground
state.)

In this paper we will show that for a given (finite) temperature a
statistically consistent microcanonical ensemble can be defined, such
that when the CBS are orthonormalized using a combination of
democratic and sequential orthonormalization methods, the diagonal
matrix elements of the Hamiltonian, $H$, are left unchanged for the
many body states in the microcanonical ensemble and the
quasiparticle-quasihole excitations from the microcanonical
ensemble. This means that these matrix elements can be evaluated by
borrowing methods directly from the zero temperature theory.

As mentioned earlier, this work can be considered to be an extension of the
orthonormalization scheme of of Ref.~\cite{FP} to finite
temperatures. However, it serves a more general purpose of introducing
a variational theory at finite temperatures which has the same
simplicity of formulation and efficiency in calculation as the corresponding zero
temperature theory.

\section{Variational theory in a Microcanonical Ensemble}

 In the previous section we have defined the non interacting  Fermi
 gas states $|\Phi_I\ran$ and the CBS $|\Psi_I)$. Let us call the
 corresponding OCBS $|\Psi_I\ran$. Note that the actual definition of
 $|\Psi_I\ran$ will depend on how we choose to orthonormalize the
 CBS. We will denote any of these OCBS by $|\Psi_I\ran$ and  the actual
 orthogonalization procedure used to obtain them will hopefully be obvious from the
 context. 
Let us also define a `microcanonical' subset $\calM (T)$, from  the set of all labels $I$
of the many body states (CBS, OCBS or non interacting Fermi gas) previously defined. We will call this set the `microcanonical
 ensemble' at temperature $T$. Henceforth the argument $T$ will be
 supressed for brevity. Note that as yet we have not really said
 anything about which elements are included.  We tackle this slightly
 non trivial problem in detail later in this section. For now, we
 assume that $\calM$ is a suitably defined `microcanonical' ensemble
 at the given temperature.
  We can legitimately define a density matrix,
\beq
\rho_{\textrm{MC}} = \frac{1}{\calN_{\calM}} \sum_{I \in \calM} |\Psi_I\ran\lan \Psi_I|~,
\eeq 
where $\calN_{\calM}$ is the number of elements in the set $\calM$. 

It is well known in statistical mechanics that the thermodynamic averages of the densities of extensive
quantities are the same in all ensembles; grand canonical, canonical or
microcanonical \cite{Fowler}. In Eq.~(\ref{freee}) we have used the canonical ensemble
for the average value of $ H $.

 With the microcanonical ensemble we obtain a simpler expression,
\beq
\label{hmc}
\lan H \ran = \frac{1}{\calN_{\calM}} \sum_{I~\in~\calM} \lan \Psi_I |H|\Psi_I \ran~.
\eeq

In order to develop the variational theory of hot quantum liquids we have to
orthonormalize the CBS in the microcanonical ensemble, $\calM$.  This can be easily achieved with the L\"{o}wdin
transformation \cite{Lowdin},  
\ba
\label{lowdin}
|\Psi_I\ran  & = & |\Psi_I)-\frac{1}{2}\sum_{J~\in~\calM}|\Psi_J)\overline{(\Psi_J|\Psi_I)}\nonumber \\
& + & \frac{3}{8}\sum_{J,K~\in ~\calM}|\Psi_K)\overline{(\Psi_K|\Psi_J)}~\overline{(\Psi_J|\Psi_I)}+\cdots~.
\ea
The coefficients $1$, $-\frac{1}{2}$,
$\frac{3}{8}$, $\cdots$ that occur in the L$\ddot{\textrm{o}}$wdin
transformation are those which are found in the expansion of $(1+x)^{-1/2}$.
The overhead bar signifies,
\beq
\overline{(\Psi_J|\Psi_K)}=(\Psi_J|\Psi_K)(1-\delta_{JK})~.
\eeq

The orthonormal states $|\Psi_I\ran$ are in one to one correspondence with the CBS $|\Psi_I)$ and
the Fermi-gas states $|\Phi_I\ran$, and we are then justified in
defining a variational  Hamiltonian $H_V$ such that
\ba
H_V |\Psi_I\{n_I(\bfk,\sigma_z)\} \ran &=& \left[ \sum_{\bfk,\sigma_z} n_I(\bfk,\sigma_z) \epsilon_V(\bfk,\sigma_z) \right] |\Psi_I\{n_I(\bfk,\sigma_z)\} \ran~,\\
                                       &=& E^V_I|\Psi_I\{n_I(\bfk,\sigma_z)\} \ran
\ea
thus removing the approximation inherent in Eq.~(\ref{nonoest}).  The CBS  $\not\in \calM$ are
not orthonormalized by the transformation (\ref{lowdin}).  Most of these states have little effect
on the thermodynamic properties of the liquid in equilibrium at temperature $T$ and density $\rho$.  Formally these states should
be first orthonormalized to those $\in \calM$ by Gram-Schmidt's method, and then orthonormalized
with each other using combinations of Gram-Schmidt and L\"{o}wdin methods
\cite{FP}.  This way their orthonormalization will have  no effect
on the states $\in \calM$. In the next section we will have occasion to discuss the
orthogonalization of a subset of these states, viz. states with one (quasi)particle
and one (quasi)hole with respect to the states in the microcanonical ensemble.  

In the variational estimate of the free energy [Eq.~(\ref{thvar})] we should use the OCBS
 rather than the CBS.  In the remaining part of this section we show that
\beq
\label{diagel}
\frac{1}{N} \lan \Psi_I |H| \Psi_I \ran = \frac{1}{N} ( \Psi_I |H|\Psi_I ) = \frac{1}{N}E_I~,
\eeq

{\it i.e.} if we define

\beq
\delta E_I = \left[\lan \Psi_I |H| \Psi_I \ran -  ( \Psi_I |H|\Psi_I )\right]
\eeq
and
\beq
\delta e_I = \frac{1}{N} \delta E_I
\eeq
then,
\beq
\delta e_I = 0
\eeq
for $I \in \calM$, in the limit $N \rta \infty$.  Therefore the variational free energy calculated
with the SP scheme does not have any orthogonality corrections.
 
At this point it is necessary to define the microcanonical ensemble ($\calM$) more carefully. Typically, a microcanonical ensemble is defined as 
\beq
\label{defmc}
\calM \equiv \bigcup\quad \textrm{ All states } I \quad \textrm{
  with } E_{\textrm{MC}} \leq E^V_I \leq E_{\textrm{MC}} + \delta E_{\textrm{MC}}~.
\eeq 
The actual value of $\delta E_{\textrm{MC}}$ is unimportant as long as
  $\delta E_{\textrm{MC}} \ll E_{\textrm{MC}}$. 
In our case it proves necessary that it takes a nonzero  value. 
In the next subsection we will show that the simplest definition of  $\calM$, 
i.e. with $\delta E_{\textrm{MC}} = 0$, gives a divergent expression  for the 
diagonal matrix elements of $H$. This exercise  will nevertheless help to  illustrate some 
of the simplest elements of the calculations that follow and will  serve as a motivation 
for the following subsection where we formulate the problem slightly  differently, which 
makes calculations more convenient, but is similar to 
  defining $\calM$ with a nonzero $\delta E_{\textrm{MC}}$.

\subsection{Energy Conserving Microcanonical Ensemble}

 Let us define a set $\calM_0$, which we will call the Energy Conserving Microcanonical Ensemble (ECMC), as the set of all states with
 
 \beq
 \label{ecmc}
  E^V_I =E_{MC} 
\eeq
Consider a many body state $I \in \calM_0$ with,
\beq
E_I = (\Psi_I|H|\Psi_I)~.\\
\eeq
Then the change in the diagonal matrix elements of the Hamiltonian due to L\"{o}wdin orthonormalization is given
by
\ba
\label{deltae}
\delta E_I  &=& \lan \Psi_I|H-E_I|\Psi_I\ran \nonumber \\
&=&
- \frac{1}{2}\sum_{J~\in~\calM_0}\left[\overline{(\Psi_I|H-E_I|\Psi_J)}~\overline{(\Psi_J|\Psi_I)}
+ \overline{(\Psi_J|H-E_I|\Psi_I)}~\overline{(\Psi_I|\Psi_J)} \right]+ \cdots ~,
\ea
where the dots denote higher order terms which can easily be obtained from Eq.~(\ref{lowdin}).  The
nondiagonal CBS matrix elements $\overline{(\Psi_J|H-E_I|\Psi_I)}$ and $\overline{(\Psi_I|\Psi_J)}$
can be evaluated with cluster expansions \cite{FP}.  The leading two-body clusters
contribute to the nondiagonal matrix elements only when two quasi-particles in $J$
are different from those in $I$.  
Let quasiparticle states with  momenta  $\mathbf{k_1}$ and $\mathbf{k_2}$ be occupied in $I$
and unoccupied in $J$, while states $\mathbf{k_{1'}}$ and $\mathbf{k_{2'}}$ be occupied in $J$ but not in
$I$. 

We will denote the CBS $|\Psi_J)$ and the OCBS $|\Psi_J\ran$ by 
\ba
|\Psi_J) \equiv |\mathbf {k_{1'},k_{2'}}: I-\mathbf{k_{1},k_{2}})~, \\
|\Psi_J\ran \equiv |\mathbf {k_{1'},k_{2'}}: I-\mathbf{k_{1},k_{2}}\ran~.
\ea

Note that in this notation $|\Psi_I)$ can  be written as $ |\mathbf {k_{1},k_{2}}: I-\mathbf{k_{1},k_{2}})$ 

The $I \rta J$ transition occurs via the scattering of two quasi-particles from states $(\mathbf{k_{1},k_{2}}) \rta (\mathbf{k_{1'},k_{2'}})$.  
Momentum conservation implies
\beq
\label{momcon}
\bfk_1+\bfk_2 = \bfk_{1'}+\bfk_{2'}~.
\eeq
Unless this condition is satisfied, the nondiagonal CBS matrix elements are zero.  

The two-body cluster contributions to $\overline{(\Psi_J|H-E_I|\Psi_I)}$ and $\overline{(\Psi_I|\Psi_J)}$
are respectively given by
\beq
\label{tbme}
\lan \mathbf{k_{1'},k_{2'} - k_{2'},k_{1'}}|v^{eff}_{ij}|\mathbf{k_1,k_2}\ran~\textrm{ and }~\lan \mathbf{k_{1'},k_{2'} - k_{2'},k_{1'}} |(\calF^2_{ij}-1)|\mathbf{k_1,k_2}\ran~,
\eeq
where the two-body effective interaction is given by:  
\beq
v^{eff}_{ij}=\calF_{ij} \left[ v_{ij} \calF_{ij} -\frac{\hbar^2}{m}(\nabla^2 \calF_{ij}
+ 2 \boldnabla \calF_{ij} \cdot \boldnabla) \right]~,
\eeq
the bare two-body interaction is denoted by $v_{ij}$, and the non
interacting  two-particle states are
\beq
\lan \mathbf{r_i,r_j}|\mathbf{k_1,k_2}\ran = \frac{1}{\Omega} e^{i(\bfk_1 \cdot \bfr_i + \bfk_2 \cdot \bfr_j)}~.
\eeq
The factor $1/\Omega$ comes from the normalization of the plane waves. We have suppressed the spin wave functions for brevity.

\begin{figure}[tbp]
\subfigure[]{\includegraphics[width = 0.4\textwidth]{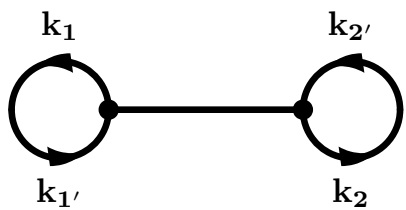}}
\hfill
\subfigure[]{\includegraphics[width = 0.4\textwidth]{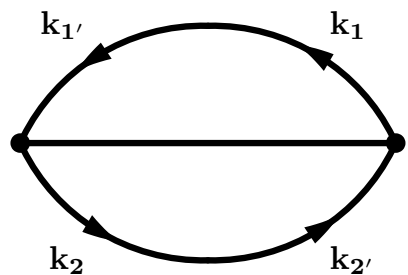}}
\\
\subfigure[]{\includegraphics[width = 0.4\textwidth]{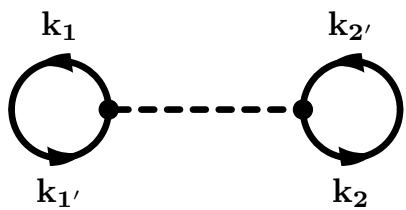}}
\hfill
\subfigure[]{\includegraphics[width = 0.4\textwidth]{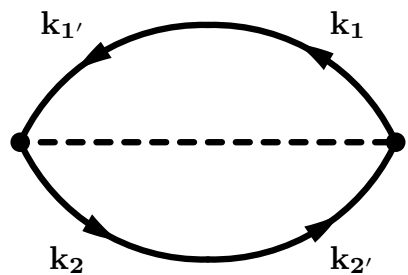}}
\caption{All two body cluster diagrams}
\label{fig1}
\end{figure}

\begin{figure}[htbp]
\subfigure[]{\includegraphics[width = 0.4\textwidth]{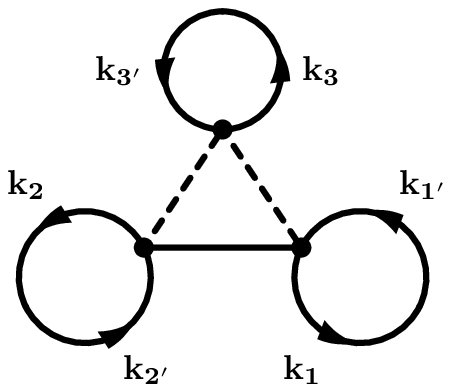}}
\hfill
\subfigure[]{\includegraphics[width = 0.4\textwidth]{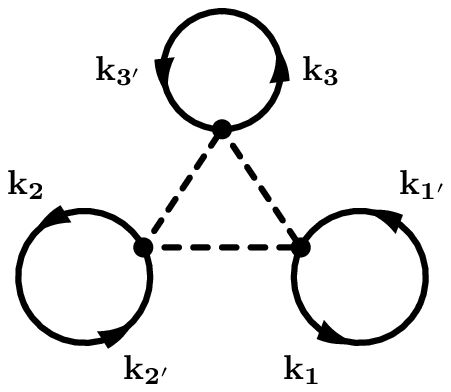}}
\caption{Examples of three body connected diagrams}
\label{fig2}
\end{figure}

\begin{figure}[htbp]
\subfigure[]{\includegraphics[width = 0.4\textwidth]{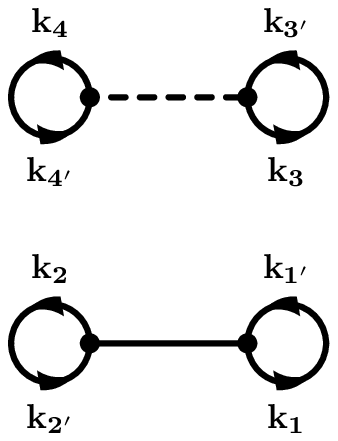}}
\hfill
\subfigure[]{\includegraphics[width = 0.4\textwidth]{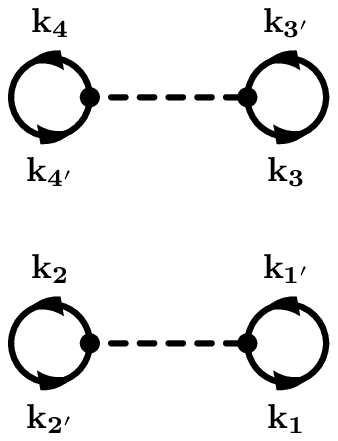}}
\caption{Examples of  disconnected diagrams}
\label{fig3}
\end{figure}

The CBS matrix elements can be represented by diagrams, such as those
in Figs.~\ref{fig1}, \ref{fig2} and \ref{fig3},
which have been analyzed in detail in \cite{FP}. We will adopt their
notation and use their results. In all the diagrams
we use the following conventions.
\bit
\item
The points in these diagrams denote positions of the particles: $\bfr_i,~\bfr_j,...$.  
\item
The dashed lines connecting points $i$ and $j$ represent correlations, i.e. terms originating from
$\calF^2_{ij}-1$.  When $\calF_{ij}=f(r_{ij})$ this notation is sufficient, however, a more elaborate
notation for the correlation lines is needed when $\calF$ is an operator with many terms \cite{PWRMP}.
For brevity we will show diagrams assuming $\calF_{ij}=f(r_{ij})$, commonly called the
Jastrow correlation function.
\item
The solid lines represent $v^{eff}_{ij}$.  There can only be one solid line in a diagram
representing matrix elements of $H$.
\item
The lines with one or two arrowheads represent state lines. The arrowheads are labeled with
quasi-particle states.  A state line with a single arrowhead labeled $\bfk_\ell$
going from point $i$ to point $j$ indicates that the particle $i$ is in state $\bfk_\ell$ in the
ket $|\Psi_I)$ and particle $j$ is in $\bfk_\ell$ in the bra $(\Psi_J|$.  Diagrams representing
diagonal CBS matrix elements can have state lines with only one arrowhead, since
the state $\bfk_\ell$ is occupied (or unoccupied) in both the bra and the ket.

Diagrams contributing to the nondiagonal CBS matrix elements have state lines with two
arrowheads.  The number of these lines equals the number of quasi-particle states that are
different in $(\Psi_J|$ and $|\Psi_I)$.  A state line with arrowheads $\bfk_\ell$ and $\bfk_{\ell'}$,
going from $i$ to $j$ indicates that $i$ is in state $\bfk_\ell$ in the ket, while $j$ is in
state $\bfk_{\ell'}$ in the bra, and that $\bfk_\ell$ and $\bfk_{\ell'}$ are unoccupied in the bra and the
ket respectively.  

Only one state line must emerge from a point and only one must end in a point because each
particle occupies only one quasi-particle state in the bra and the ket. This implies that
the state lines form continuous loops.  In direct diagrams the state lines emerge from and end
on the same particle, while in exchange diagrams they connect pairs of particles.
\item
The contribution of a diagram is given by an integral over all the particle coordinates
$\bfr_i$ in the diagram.  The integrand contains factors of $(f^2(r_{ij})-1)$ for each
correlation line, $v^{eff}_{ij}$ for the interaction line, $e^{i\bfk_{\ell}
\cdot \bfr_i}/\sqrt{\Omega}$ for each state line $\ell$ emerging from
a point $\bfr_i$
and $e^{-i\bfk_{\ell'}\cdot\bfr_j}/\sqrt{\Omega}$ for each state line $\ell'$ ending in $\bfr_j$.  
\eit

The two-body ($2b$) direct ($d$) and exchange diagrams ($e$) representing the nondiagonal matrix elements (\ref{tbme})
are shown in Fig.~\ref{fig1}.  The contributions of the $2b.v^{eff}$ diagrams
are given by

\ba
 2b.v^{eff}.d=\frac{1}{\Omega} \int d^3r_{ij} \exp(-i\frac{1}{2}(\bfk_{1'}-\bfk_{2'})\cdot\bfr_{ij})
                       v^{eff}(r)\exp(i\frac{1}{2}(\bfk_1-\bfk_2)\cdot\bfr_{ij})~,\\
 2b.v^{eff}.e= -\frac{1}{\Omega} \int d^3r_{ij} \exp(-i\frac{1}{2}(\bfk_{2'}-\bfk_{1'})\cdot\bfr_{ij})
                       v^{eff}(r)\exp(i\frac{1}{2}(\bfk_1-\bfk_2)\cdot\bfr_{ij})~,
\ea
 
for spin independent $v$ (and $f$) .  The contributions of  $2b.\calF^2$ diagrams
 are obtained by replacing $v^{eff}$ by $\calF^2-1$.  All $2b$ diagrams have a contribution of
 order $1/\Omega$.

   The change in energy, $\delta E_I$, [Eq.~(\ref{deltae})] contains products of the $2b.v^{eff}$ and $2b.\calF^2$
diagrams.  These products are of order $1/\Omega^2$.  The total contribution of the leading $2b$
cluster terms to $\delta E_I$ is obtained by summing over the states $\mathbf{k_{1},k_{2},k_{1'},k_{2'}}$.  Each
allowed combination of these states corresponds to a many-body state in the set $\calM_0$.
The quasi-particle states $\mathbf{k_{1}}$ and $\mathbf{k_2}$ can be any two of those occupied in $I$.  Thus
the sum over these gives a factor of order $N^2$.  Next we sum over $\mathbf{k_{1'}}$ and $\mathbf{k_{2'}}$.  The total momentum $\bfk_{1'}+\bfk_{2'}$ is
determined from Eq.~(\ref{momcon}).  The magnitude of the relative momentum,
\beq
\bfk_{1'2'}=\frac{1}{2}(\bfk_1'-\bfk_2')~,
\eeq
is constrained by energy conservation,
\beq
\epsilon_V(\bfk_1)+\epsilon_V(\bfk_2)=\epsilon_V(\bfk_{1'})+\epsilon_V(\bfk_{2'})~.
\eeq
required for $J$ to be in the set $\calM_0$.  
Thus the sum over states allowed for $\mathbf{k_{1'}}$ and $\mathbf{k_{2'}}$ corresponds to an integration over the direction of $\bfk_{1'2'}$.
It gives a factor of order $\Omega^{1/3}$. Hence
\ba
\delta e_I.2b &=& \frac{1}{N} \delta E_I.2b\\  
& \sim & \frac{1}{N}\frac{1}{\Omega ^2} N^2 \Omega ^{1/3}\\ 
& \sim & \rho \Omega ^{-2/3}  
\ea

{\it i.e.} $ \delta e_I.2b\rta 0$ as $N \rta \infty$.  Note that with the constraint of momentum conservation alone we can integrate
over the magnitude of $\bfk_{1'2'}$.  This integration gives a factor of order $\Omega$ and makes $\delta e_I.2b$
of order $1$.  The equal energy constraint  in the ECMC $\calM_0$ (
Eq.~(\ref{ecmc})) makes $\delta e_I.2b$ vanish in the thermodynamic limit.

The above analysis can be carried out for contribution of clusters with three or more particles to
$\delta E_I$.  Consider, for example, states $J$ which differ from $I$ in occupation numbers of three
quasi particles.  These states can be reached by scattering three quasi-particles in $I$ in states
$\bfk_1,\bfk_2,\bfk_3$ to states $\bfk_{1'},\bfk_{2'},\bfk_{3'}$ occupied in $J$.  The relevant, direct $3b$ diagrams are shown in Fig.~\ref{fig2}.  
Each diagram is  
of order $1/\Omega^2$. The  contribution of three body cluster
terms to $\delta E_I$ are products of $v^{eff}$ and $\calF^2 -1$
diagrams. Each of these products  is of order $1/\Omega^4$.  We get a
factor of order $N^3$ by summing over states $\bfk_1,\bfk_2$ and $\bfk_3$, and a factor $\Omega^{4/3}$ by summing over $\mathbf{k_{1'},k_{2'},k_{3'}}$ with
constraints of momentum and energy conservation.  Thus their total
contribution to $\delta e_I$ is of order $N^{-1}(\Omega^{-4} N^3
\Omega ^{4/3}) \sim \rho^2 \Omega^{-2/3}$ which vanishes in the
thermodynamic limit just like the contribution from the leading two
body cluster terms. Similarly the contribution from all connected
terms can be shown to give vanishing contribution to $\delta e_I$ in
the thermodynamic limit.

The terms of Eq.~(\ref{deltae}) will also contain disconnected diagrams
like those shown in Fig.~\ref{fig3}. These diagrams by themselves  will give rise to unphysical divergent 
(non extensive) contributions to the energy. To extract any physically
meaningful result from the theory, these diagrams must cancel identically.

 Disconnected diagrams in the expansion for the shift in energy ($\delta E_I$) can be classified into two types.
\begin{itemize}
 \item Diagrams in which each connected cluster conserves energy. 
 \item Diagrams in which only the whole diagram conserves energy, i.e. each connected 
    cluster does not conserve energy.
\end{itemize}
 Let us, consider the first case. 

Consider the simplest possible divergent diagrams, i.e. Fig.~\ref{fig3}.  

Let us denote the CBS $|\Psi_J)$, $|\Psi_K)$ and $|\Psi_L)$ by
\ba
\label{dis}
|\Psi_J)&=&|\mathbf{k_{1'},k_{2'},k_{3'},k_{4'}} : I - \mathbf{k_1,k_2,k_3,k_4}) \nonumber \\
|\Psi_K)&=&|\mathbf{k_{1'},k_{2'}} : I - \mathbf{k_1,k_2}) \nonumber \\
|\Psi_L)&=&|\mathbf{k_{3'},k_{4'}} : I - \mathbf{k_3,k_4})
\ea

Let $\mathbf{k_1,k_2}$ and $\mathbf{k_{1'},k_{2'}}$ conserve momentum \emph{and energy} amongst themselves and
similarly for $\mathbf{k_3,k_4}$ and $\mathbf{k_{3'},k_{4'}}$. 

\ba
\bfk_1+\bfk_2 = \bfk_{1'}+\bfk_{2'}~,\label{mom1}\\
\bfk_3+\bfk_4 = \bfk_{3'}+\bfk_{4'}~, \label{mom2}
\ea

and

\ba
\epsilon_V(\bfk_1)+\epsilon_V(\bfk_2) = \epsilon_V(\bfk_{1'})+\epsilon_V(\bfk_{2'})~,\label{eng1}\\
\epsilon_V(\bfk_3)+\epsilon_V(\bfk_4) = \epsilon_V(\bfk_{3'})+\epsilon_V(\bfk_{4'})~.\label{eng2}
\ea
An inspection of the series on the right hand side of Eq.~(\ref{deltae}) will show that
thirteen (13) terms in total will give rise to an integral which
will be represented by products of diagrams shown in Fig.~\ref{fig3}. In
Table~\ref{contrb} we list the terms 
and also their corresponding prefactor in the series. As one can see, the sum of the prefactors is identically
zero. Thus products of diagrams of the type shown in Fig.~\ref{fig3} has
no contribution to the change in energy
per particle ($\delta e_I$). The cancellation of corresponding
exchange diagrams and all other divergent diagrams of this order with
three or more body connected pieces can also be shown to cancel with
analogous book keeping.  The divergent diagrams
of the next highest order can also be shown to cancel identically.

  Now consider the case when the individual clusters \emph {do not} conserve energy, i.e. 
Eqs.~(\ref{mom1},~\ref{mom2}) are still true but Eqs.~(\ref{eng1},~\ref{eng2}) are \emph{not} true. Instead,
\beq
\epsilon_V(\bfk_1)+\epsilon_V(\bfk_2)+\epsilon_V(\bfk_3)+\epsilon_V(\bfk_4) = \epsilon_V(\bfk_{1'})+\epsilon_V(\bfk_{2'})+ \epsilon_V(\bfk_{3'})+\epsilon_V(\bfk_{4'})~.\label{eng}
\eeq

 In this case the states $K$ and  $L$ no longer belong to the same ECMC as $I$ and  $J$. Thus, none of the
 terms in Table 1, except for  the first, will be included in the sum, i.e. the divergent terms,
 $\left(\Psi_I\right|v^{eff}\left|\Psi_J\right)\left(\Psi_J\right.\left|\Psi_I\right)$ and
 its complex conjugate will not get cancelled. The total contribution of
 terms like these to $\delta E_I$ is of order $\rho^4 \Omega^{4/3}$, i.e. the
 shift in the energy per particle, $\delta e_I \sim \rho^3 \Omega^{1/3}$, diverges
 in the thermodynamic limit. 
  
  The survival of divergent terms is rather artificial and arises from
  the sharp energy conservation constraint that we imposed on the states in $\calM_0$. 
This provides the motivation to define a ensemble where this constraint is relaxed slightly. 
In what follows we will show that this can be done consistently where  none of the divergent 
terms are present while the diagonal matrix elements of the Hamiltonian are preserved.

\subsection{The set of Most Probable Distributions (MPD)}

 Consider an ideal gas of fermions in a box of volume
 $L^3(= \Omega)$ . The single particle energy levels are given by
\beq
\label{sps}
\epsilon_V(\mathbf {n}_i) = \frac{\hbar^2}{2m}\frac{{\mathbf n}_i^2}{2\pi L^2}  
\eeq
 where $m$ is the mass of the fermions and ${\mathbf n}_i$ is a vector
 with integer components. In what follows we will use units where
 $\frac{\hbar^2}{4\pi m} = 1$. Let the total number of
 particles in the box be $N$. 

The density of states at any single particle energy
 $\epsilon_V$ is given by
\beq
g(\epsilon_V) \sim L^3 \epsilon_V^{1/2}
\eeq

Now consider a cell with an energy width of $\frac{\Delta}{L^2}$, around an energy
level $\epsilon_V$ . Then the number of single particle energy levels in
this cell is
\beq
\omega(\epsilon_V)\sim g(\epsilon_V)\frac{\Delta}{L^2} \sim L^3 \epsilon_V^{1/2}\frac{\Delta}{L^2}~.
\eeq 
 The exact value of $\Delta$ is not important, except for the fact
 that it is dimensionless and of order 1 or less. We can always choose $\Delta$  so that $\omega(\epsilon_V)$ is large,
\beq
\omega(\epsilon_V) \gg 1~.
\eeq

  Let the total number of particles in the energy cell around $\epsilon_V$ be $n(\epsilon_V)$.    
 Let $S({n(\epsilon_V)})$ be the entropy corresponding to the configuration
(distribution) $n(\epsilon_V)$. It can be easily shown that the
 entropy, $S_V$ corresponding to the \emph{most probable distribution} of number of
 particles per \emph {energy cell} ,
 subject to the constraints 
\ba
\sum n(\epsilon_V) &=& N \label{cons1}\\
\textrm{and} \qquad \sum \epsilon_V n(\epsilon_V) &=& E_{MC}\label{cons2}~,
\ea
 is given by Eq.~(\ref{ent}), {\it i.e.}, the distribution
 $\overline{n}(\epsilon_V)$ in addition to satisfying
 Eqs.~(\ref{cons1},\ref{cons2}) also satisfies the maximization
 condition,
\beq
S(\overline{n}(\epsilon_V)) = \textrm{Maximum} (S(n(\epsilon_V))) 
\eeq
The temperature and the chemical potential are the Lagrange multipliers of the minimization procedure.  
  Now we will define our microcanonical ensemble as the set $\calM$ of all configurations
 whose cell distribution is ${\overline{n}(\epsilon_V)}$. We will call this the set of Most Probable Distributions (MPD). Please note that this is  different from defining an ensemble with total
 energy $E_{\textrm{MC}}$, but is roughly the same as defining an ensemble with an
 average energy $E_{\textrm{MC}}$ and a small non zero energy width $\delta E_{\textrm{MC}}$.
\beq
\calM \equiv \bigcup (\textrm { All states with the distribution }{\overline{n}(\epsilon_V)}  )
\eeq
 It should be emphasized here that none of the conclusions that follow
 depend explicitly  on the actual single particle spectrum given by Eq.~(\ref{sps}). 
All the conclusions remain unchanged as long as the single particle energy
 levels produce a continuum in the thermodynamic limit. However we
 will continue to use Eq.~(\ref{sps}) because the calculations are more transparent this way.

The probability of fluctuations about the most probable distribution
is given by Einstein's relation,
\beq
P(\delta n) \sim e^ {- \delta S}~,
\eeq 
where  $P(\delta n)$ is the probability of a fluctuation of size
$\delta n$ and $\delta S$ is the corresponding \emph {decrease} in
entropy. Around the most probable distribution,
\beq
\delta S \sim \delta n^2~,
\eeq
i.e. the probability of fluctuations vanishes exponentially with the
size of the fluctuations.

In addition, the fluctuation (standard deviation) in the value of the 
total energy, $E^V_I$, in $\calM$, can be easily shown to be,
\beq
\delta E < \frac{\Delta}{L^2} \sqrt{N} 
\eeq
i.e. $\delta E$ is \emph {non} macroscopic; the fluctuation in the
energy per particle vanishes in the thermodynamic limit. Thus, 
 the ensemble we have defined is a consistent one in the
statistical sense.

  Now let us discuss the allowed scattering processes within the set
 $\calM $. For two states to be in $\calM$, they must have the same
 populations $n(\epsilon_V)$ in all the energy bins (cells). What this means is that they must be
 connected to each other through excitations \emph {within}
 cells. For example let $I $ and $\mathbf{k_{1'},k_{2'}}:I - \mathbf {k_1,k_2}$, be elements of
 $\calM$. Let
 \ba
 \epsilon_V(\mathbf{k_{1}})&\leq&\epsilon_V(\mathbf{k_2})\\
 \textrm{and} \qquad\epsilon_V(\mathbf{k_{1'}})&\leq&\epsilon_V(\mathbf{k_{2'}})~.
 \ea
Then, we need to have
\ba
 |\epsilon_V(\mathbf{k_{1'}}) - \epsilon_V(\mathbf{k_1})| &<& \frac{\Delta_1}{L^2} \label{appeng1}\\ 
 \textrm{and} \qquad |\epsilon_V(\mathbf{k_{2'}}) - \epsilon_V(\mathbf{k_2}) |&<&\frac{\Delta_2}{L^2} \label{appeng2}
\ea
where $\frac{\Delta_1}{L^2}$ and $\frac{\Delta_2}{L^2}$ are the widths of the cells containing
$\mathbf{k_1}$ and $\mathbf{k_2}$ respectively. Please note that this, in general, means that there is no exact energy conservation, but that there is approximate energy conservation for each individual particle. 

Let us discuss the orthogonalization correction in $\calM$. Consider
the \emph{disconnected} diagrams first. 
Following the discussion before, let us define states  $J$, $K$ and
$L$ as in Eq.~(\ref{dis}), with,
\ba
\epsilon_V(\mathbf{k_1})\leq \epsilon_V(\mathbf{k_2})\leq \epsilon_V(\mathbf{k_3})\leq \epsilon_V(\mathbf{k_4})~, \nonumber\\
\epsilon_V(\mathbf{k_{1'}})\leq \epsilon_V(\mathbf{k_{2'}})\leq\epsilon_V(\mathbf{k_{3'}})\leq \epsilon_V(\mathbf {k_{4'}})~. 
\ea
Let us assume that they obey Eqs.~(\ref{mom1},~\ref{mom2}) i.e. they form two disconnected momentum conserving clusters. 

Now it is crucial to observe that if $I$ and $J$ belong to $\calM$
then $\mathbf{k_1}$ and $\mathbf{k_{1'}}$ must belong to the same energy cell; similarly for $\bfk_{2}$
and $\bfk_{2'}$, $\bfk_3$ and $\bfk_{3'}$ , and $\bfk_4$ and $\bfk_{4'}$.  This implies
that $K$ and $L$ must also belong to $\calM$. This was not the case
 when we had merely imposed overall exact energy conservation.
 
 Thus, in this case all the terms in Table~\ref{contrb} will contribute to the
 sum in Eq.~(\ref{deltae}). As such the disconnected diagrams will
 cancel each other and we will be left with a connected, non divergent
 sum, {\it i.e.}
\beq
\delta E^{\textrm{disconnected}}_I = 0
\eeq
identically. The problem with disconnected diagrams that we encounter in ECMC is resolved in MPD.
 
 For the sake of completeness, we show that the connected diagrams also have a vanishing contribution towards the energy in the thermodynamic limit.
Consider the two body cluster contributions to $\delta E_I$. 
The $\delta E_I$ [Eq.~(\ref{deltae})] contains products of the $2b.v^{eff}$ and $2b.\calF^2$
 diagrams.  These products are of order $1/\Omega^2$.  The total contribution of the leading 2b
 cluster terms to $\delta E_I$ is obtained by summing over the states
 $\mathbf{k_1}$, $\mathbf{k_2}$, $\mathbf{k_{1'}}$ and $\mathbf{k_{2'}}$.  Each
 allowed combination of these states corresponds to a many-body state in the set $\calM$.
 The quasi-particle states $\mathbf{k_1}$ and $\mathbf{k_2}$ can be any two of those occupied in $I$.  Thus,
the sum over $\mathbf{k_1}$ and $\mathbf{k_2}$ gives
 a factor of order $N^2$.  Next we sum over $\mathbf{k_{1'}}$ and $\mathbf{k_{2'}}$.  The total momentum $\bfk_{1'}+\bfk_{2'}$ is
 determined from Eq.~(\ref{momcon}).  The magnitude of the relative momentum:
 \beq
 \bfk_{1'2'}=\frac{1}{2}(\bfk_1'-\bfk_2')~,
 \eeq
 is constrained by Eqs.~(\ref{appeng1},~\ref{appeng2}).

The sum over states allowed for $\mathbf{k_{1'}}$ and $\mathbf{k_{2'}}$ corresponds to an
 integration over  $\bfk_{1'2'}$. But $\bfk_{1'2'}$ is
 constrained to lie in a shell of width $\Delta$ where,
 $\Delta \sim \min(\Delta_1,\Delta_2)$. Thus, the sum over $\bfk_{1'}$ and $\bfk_{2'}$
gives a contribution $\sim \Omega \frac{\Delta}{L^2}$ up to a factor
 of order 1 (the factor of $\Omega$
 comes from the density of states). Therefore the total contribution of
 the 2b diagrams after summing over $\mathbf{k_1,k_2,k_{1'},k_{2'}}$  is,
\beq
\delta E_I.2b\sim \Omega \rho^2\frac{\Delta}{L^2}~.
\eeq

Thus the shift in the energy per particle is,
\beq
\delta e_I.2b \sim \rho \frac{\Delta}{L^2}~,
\eeq 
which vanishes in the thermodynamic limit.

The above analysis can be easily carried out for contribution of
\emph{ connected}
clusters with three or more particles to the
$\delta E_I$.  Consider for example states $J$ which differ from $I$ in occupation numbers of three
quasi particles.  These states can be reached by scattering three quasi-particles in $I$ in states
$\mathbf{k_1,k_2,k_3}$ to states $\mathbf{k_{1'},k_{2'},k_{3'}}$ occupied in $J$. For example consider the direct 3b term shown in Fig.~\ref{fig2} .  Each is  
of order $1/\Omega^2$, thus their contribution to $\delta E_I$ is of order $1/\Omega^4$.  We get a
factor of order $N^3$ by summing over $\mathbf{k_1,k_2,k_3}$, and a factor
$\Omega^2 \left(\frac{\Delta}{L^2}\right)^2$ by summing over $\mathbf{k_{1'},k_{2'},k_{3'}}$ with
constraints of momentum and (approximate) energy conservation.  Thus
their total is of order $\Omega \rho^3 \left(\frac{\Delta}{L^2}\right)^2 $. The contribution to the shift in
energy per particle is $\delta e_I \sim \rho^2
\left(\frac{\Delta}{L^2}\right)^2 $ which vanishes in the
thermodynamic limit: similarly for higher order clusters. 
Therefore, for connected clusters we see that
\beq
\label{connected}
\delta e_I^{\textrm{connected}} \rightarrow 0~, 
\eeq 
in the thermodynamic limit.

  Thus, as claimed earlier in the section we have shown that it is
  possible to define a statistically consistent microcanonical
  ensemble such that Eq.~(\ref{diagel}) is true for the elements in
  the microcanonical ensemble $\calM$.

\subsection{Discussion}

 The simplest choice for a microcanonical ensemble is the ECMC. The ECMC has the following properties:
\begin{itemize}
\item The states in ECMC have exact energy conservation; this imposes a sharp energy cutoff.
\item Arbitrarily high single particle energy transfers are allowed while still remaining in the same ECMC 
\end{itemize} 

 It is due to the second property that clusters in disconnected
 diagrams can have arbitrary energy transfers and hence divergent
 contributions. These contributions are normally (in a canonical
 ensemble, when all states are included) canceled by contributions
 from higher order terms. But by imposing exact energy conservation
 we exclude the states which lead to these higher order terms which
 cancel the divergent part. Thus, we are left with a divergent series. 
  
In  MPD on the one hand we relax the energy conservation slightly, 
and on the other hand we limit single particle energy transfers to the width of the energy cells. 
We showed that this leads to a convergent series. Also, we showed that
the total energy is well defined in a MPD and that states with large
deviations from the MPD (i.e. states with large single particle energy transfers) are exponentially improbable.

  The main difference between ECMC and MPD is that one follows from a
  conservation law and the other from a distribution. There can be
  states in the ECMC whose population distribution in the energy cells
  is very different from the MPD $\overline {n}(\epsilon_V)$, but as
  long as the total energy of the state, $E_V = E_{MC}$, this state is
  a valid member of the ECMC. However, the number of these states is
  negligible as compared to the total number of states which have the
  MPD, and hence they can be safely neglected in the thermodynamic
  limit. On the other hand, those states  whose total energy $E_V$
  differ from $E_{MC}$ by a non macroscopic amount and have the same population
  distribution as the MPD should be included in the microcanonical ensemble. We have
  shown that, for our purposes, a typical state in a microcanonical ensemble is 
given by the (most probable) distribution and not by an exact conservation law.

Thus,  we have shown that a consistent choice for $\calM$ does exist. We have
also shown that the most obvious choice, namely, the ECMC leads to
divergences in the theory. We traced these divergences to the
existence of the sharp cutoff due to the exact energy conservation 
imposed on the states. Then we showed that these unphysical
divergences can be removed by relaxing the energy conservation
slightly, with the set of MPDs. We showed that MPDs can be
consistently treated as microcanonical ensembles and that the diagonal
matrix elements of the Hamiltonian remain unchanged upon
orthogonalization in this ensemble.

  In practical calculations, a microcanonical sample, $\calM$ of CBS is given by
\ba
\label{psimc1}
|\Psi_{MC}) &=& |\Psi\{ n_{MC}(\bfk,\sigma_z)\})~,\\
\label{psimc2}
n_{MC}(\bfk,\sigma_z) &=& 1~\textrm {with  probability}~\barn(\bfk,\sigma_z);~\textrm {else zero}~.
\ea

The state $|\Psi_{MC})$ (Eq.~\ref{psimc1}) belongs to the MC ensemble with energy
\beq
\label{emc}
E_{V,MC}=\sum_{\sigma_z} \int \frac{d^3k}{(2\pi)^3} n_{MC}(\bfk,\sigma_z) \epsilon_V(\bfk,\sigma_z)~.
\eeq
Since the Hamiltonian $H_V$ can be easily solved, we can find the
temperature corresponding to this energy.  In the $N \rta \infty$ limit it is just that
used to find the $\barn({\bfk,\sigma_z})$ [Eq.~(\ref{barn})].  All the MC states belonging to
this set can be found by allowing particles in $|\Psi_{MC})$ to scatter into \emph {allowed} final states. Each scattering produces a new CBS belonging
to the same MC set.  We denote this set by $\calM$.  If the quantum liquid is contained
in a thin container with negligible specific heat,  then it passes through the states in $\calM$
when in equilibrium at temperature $T$ and density $\rho$.

Eqs. (\ref{psimc1},~\ref{psimc2}) have been recently used to calculate the rates of weak interactions in hot nuclear
matter~ \cite{CP05}.  Note that $|\Psi_{MC})$ is a CBS since $n_{MC}(\bfk,\sigma_z)$ are
either 1 or 0.  When the number of particles in $|\Psi_{MC})$ is large the fluctuations
due to sampling the probability distribution $\barn(\bfk,\sigma_z)$ are negligible, and
this state has the desired densities $\rho_{\pm}$ and energy per particle appropriate for
the desired temperature $T$ and Hamiltonian $H_V$ used to calculate the $\barn$.
Neglecting these fluctuations in the limit $N \rta \infty$
we obtain the variational estimate for the free energy, 
\beq
\label{vareqfe}
F_V(\rho,T) = \textrm {minimum of}~\big[ ( \Psi_{MC} | H | \Psi_{MC} ) - T S_V(\rho,T) \big]~,
\eeq
where the minimum value is obtained by varying the $\calG$ and $\epsilon_V(\bfk,\sigma_z)$.
The $( \Psi_{MC} | H | \Psi_{MC} ) $ can be calculated with standard cluster expansion
and chain summation methods used in variational theories of cold quantum liquids
\cite{PWRMP,mpr}.  At low temperatures ($\ll T_F$) this method is particularly simple
because the zero temperature $\calG$ and $\epsilon(\bfk,\sigma_z)$ provide very good
approximations to the optimum.  The main concerns raised in past applications
\cite{FPS,FP81} of the SP scheme is that it neglects the nonorthogonality of the CBS,
and provides only upperbounds for the free energy.  Here we address only the first.
At zero temperature the difference between the variational and the exact ground state
energy has been estimated with correlated basis perturbation theory
\cite{FFP}. It  may be possible to extend these methods to finite
temperatures.

\section{Orthonormalization of the quasiparticle-quasihole excitations}
In the calculation of nuclear response functions one needs to use the diagonal
matrix elements of the Hamiltonian in the quasiparticle-quasihole  states \cite{FPRF}. At
least at zero temperature the leading contribution to the dynamic
structure function comes the 1p-1h states. Here we will limit our
discussion to the diagonal matrix elements of the Hamiltonian in the 1p-1h excitations
from the states in $\calM$.
 
Consider a OCBS $|\Psi_I\ran$, $I \in \calM$, where the single
quasiparticle state with momentum 
$\mathbf {h}$ is occupied but the single quasiparticle state  with
momentum $\mathbf{h+k}$ is not. 
We will denote the quasiparticle-quasihole OCBS where  the
quasiparticle state with momentum $\mathbf{h}$ is 
replaced by one with momentum $\mathbf{h+k}$ by $|\mathbf{h+k}: I -\mathbf{h}\ran$, 
and the corrresponding CBS by $|\mathbf{h+k}: I - \mathbf{h} )$. The
CBS $|\mathbf{h+k}: I - \mathbf{h} )$ 
is orthogonal to all the states in $\calM$, because they have
different total momenta. 
Thus it only needs to be orthonormalized with all the other
quasiparticle-quasihole excitations with the same momentum, 
via the L\"{o}wdin method. The excitations with two or more
quasiparticles and quasiholes 
should be  orthonormalized with the quasiparticle-quasihole states
using a sequential  method. 
But this does not have any effect on the quasiparticle-quasihole states, so we do not discuss them any further.

The quasiparticle-quasihole OCBS is given by
\beq
|\mathbf{h+k}:I-\mathbf{h}\ran = |\mathbf{h+k}: I-\mathbf{h}) - \frac{1}{2}\sum_{I' \in \calM} |\mathbf{h'+k}: I'-\mathbf{h'})\left(\mathbf{h'+k}: I'-\mathbf{h'}|\mathbf{h+k}: I-\mathbf{h}\right) + \cdots~,
\eeq
where $h' \in I'$. The diagonal matrix elements of $H$ are given by
\ba
\lan \mathbf{h+k}:I-\mathbf{h}|H| \mathbf{h+k}:I-\mathbf{h}\ran &=&\left(\mathbf{h+k}:I-\mathbf{h}|H| \mathbf{h+k}:I-\mathbf{h}\right) \nonumber \\
 &-& \frac{1}{2}\sum_{I' \in  \calM}\left[\left(\mathbf{h+k}:I-\mathbf{h}|H|\mathbf{h'+k}:I'-\mathbf{h'}\right ) \right .\nonumber\\
&\times&\left .\left(\mathbf{h'+k}:I'-\mathbf{h'}|\mathbf{h+k}:I-\mathbf{h}\right) + c.c.\right] + \cdots
\ea

In actual calculation of response functions one needs the difference
between $\lan \mathbf{h+k}:I-\mathbf{h}|H|\mathbf{h+k}:I-\mathbf{h}\ran$
and $\lan \Psi_I|H|\Psi_I\ran$. Let us define
\beq
\label{eph}
E_{ph} = \left [ \lan  \mathbf{h+k}:I-\mathbf{h}|H|\mathbf{h+k}:I-\mathbf{h}\ran - \lan  \Psi_I|H|\Psi_I\ran\right ] ~.
\eeq
 At zero temperature, in accordance with Landau's theory, one can define single particle energies for
  quasiparticle and quasiholes as done in Eq.~\ref{eqei}. The quantity
  $E_{ph}$ is analogous to (\emph{i.e.} is a finite temperature
  generalization of) the difference between the quasiparticle energy,
  $\epsilon_0(\mathbf{h+k})$, and the quasihole energy,
  $\epsilon_0(\mathbf{h})$. We will show that $E_{ph}$ has no
  orthogonality corrections.

 The orthogonality correction to $E_{ph}$ is given by
\ba
\label{ph}
\delta E_{ph} &=& \left [ \lan \mathbf{h+k}:I-\mathbf{h}|H|
  \mathbf{h+k}:I-\mathbf{h}\ran - \lan \Psi_I|H|\Psi_I\ran\right ] \nonumber\\          
&-& \left [ \left ( \mathbf{h+k}:I-\mathbf{h}|H| \mathbf{h+k}:I-\mathbf{h}\right ) -  \left ( \Psi_I|H|\Psi_I \right ) \right ]\\
   &=& \left [ \frac{1}{2}\sum_{I' \in \calM}\left[
  \left(\mathbf{h+k}:I-\mathbf{h}|v^{eff}|\mathbf{h'+k}:I'-\mathbf{h'}\right
  )
  \left(\mathbf{h'+k}:I'-\mathbf{h'}|\mathbf{h+k}:I-\mathbf{h}\right)
  \right . \right .\nonumber \\
&+& \left . \left . \textrm{c.c.}\right] + \cdots \right ]\nonumber \\ 
&-& \left [ \frac{1}{2}\sum_{I' \in
  \calM}\left[\left(\mathbf{h}:I-\mathbf{h}|v^{eff}|\mathbf{h'}:I'-\mathbf{h'}\right
  ) \left(\mathbf{h'}:I'-\mathbf{h'}|\mathbf{h}:I-\mathbf{h}\right)
  \right . \right . \nonumber \\ 
&+&\left . \left . \textrm{c.c.} \right] + \cdots \right ]
\ea 
 The non diagonal matrix elements of the Hamiltonian ($v^{eff}$) and unity in the second
  equation  are of order  $1/\Omega$ or less.

The shift $\delta E_{ph}$ will contain both connected and disconnected
terms. 
The disconnected terms can be shown to cancel exactly using arguments similar to the ones used in the last section. We will consider the connected terms only.

The state  $|\mathbf{h'+k}: I' - \mathbf{h'} )$ can be of the following types:
\begin{itemize}
\item Type 1 : $\mathbf{h'} = \mathbf{h}$ , $I' \neq I$
\item Type 2 : $\mathbf{h'} \neq \mathbf{h}$.
\end{itemize}

For Type 1 terms, the terms of the matrix elements $
\left(\mathbf{h+k}:I-\mathbf{h}|v^{eff}|\mathbf{h+k}:I'-\mathbf{h'}\right )$ 
and
$\left(\mathbf{h+k}:I'-\mathbf{h}|\mathbf{h+k}:I-\mathbf{h}\right)$ do
not depend on $\mathbf{k}$. 
(The contribution of the terms which contain the exchange line
$\mathbf {h+k}$ vanish in the thermodynamic limit 
as compared to the leading order terms. This can be easily seen by
explicitly writing 
down the cluster expansion for the matrix elements.) Thus the
contribution of these matrix elements 
is canceled by the corresponding terms
$\left(\mathbf{h}:I-\mathbf{h}|v^{eff}|\mathbf{h}:I'-\mathbf{h}\right )$ 
and $\left(\mathbf{h}:I'-\mathbf{h}|\mathbf{h}:I-\mathbf{h}\right)$. 

Consider the Type 2 CBS. Let $\mathbf{h_1}$ be the single
quasiparticle state in $I'$ which is in the same energy cell (as
defined in the last section) as $\mathbf{h}$, and let $\mathbf{h'_1}$
be the single quasiparticle state in $I$ which is in the same energy
cell as $\mathbf{h'}$. Since both  $I$ and $I'$ belong to $\calM$,
there will be at least one choice for $\mathbf{h_1}$ and
$\mathbf{h'_1}$, although in general their choice is not unique.  
\beq
|\mathbf{h'+k}: I' - \mathbf{h'}) \equiv|\mathbf{h'+k,h_1}: I' - \mathbf{h',h_1})~.
\eeq
 Similarly $|\mathbf{h+k}: I - \mathbf{h} )$ can be written as
 $|\mathbf{h+k,h'_1}: I - \mathbf{h,h'_1} )$. 
For Type 2 CBS the leading contribution to Eq.~(\ref{ph}) comes from
states where  
$I' - \mathbf{h',h_1} \equiv I - \mathbf{h,h'_1}$. In this case each
of the matrix elements  
$\left(\mathbf{h+k,h'_1}:I-\mathbf{h,h'_1}|v^{eff}|\mathbf{h'+k,h_1}:I'-\mathbf{h',h_1}\right)$ 
and
$\left(\mathbf{h'+k,h_1}:I'-\mathbf{h',h_1}|\mathbf{h+k,h'_1}:I-\mathbf{h,h'_1}\right)$ 
are  $1/\Omega$. Also, $\mathbf{h'_1}$ can be any of the
occupied quasiparticle states in $I$. 
Hence summing over $\mathbf{h'_1}$ gives a factor of $N$. The sum over
$\mathbf{h'}$ and $\mathbf{h_1}$  
along with momentum conservation and approximate energy conservation
gives a term of order $\Omega \frac{\Delta}{L^2}$. 
Note that there is no summation over $\mathbf{h}$. Thus the total
leading order contribution 
 from the Type 2 states is of order $\rho \frac{\Delta}{L^2}$, which vanishes in the thermodynamic limit.

Thus,
\beq
\delta E_{ph} \rightarrow 0~. 
\eeq

There is no orthogonality correction to the energy differences which enter the calculations of response functions.

\section{Conclusion}
We have developed a variational theory for hot quantum liquids. We
have shown that the correlated basis states which
provide a reasonable description of the ground state of quantum
liquids can be used to describe quantum liquids at finite
temperature. Although the correlated basis states are not orthogonal to
each other by construction, the free energy calculated in a suitably
defined microcanonical ensemble of the correlated basis does not have
any corrections due to nonorthogonality. As such the powerful cluster
expansion and chain summation methods developed for zero temperature
quantum liquids can be used at finite temperature without any
reformulation. We have also shown that the energy differences which are needed for
calculating response functions do not need any orthogonality corrections
either.

We wish to emphasize that the arguments used in this work do not
depend on the detailed form of the correlation functions or the choice
of the trial single quasiparticle spectrum. The correlated functions
are merely required to be sufficiently well behaved so that all the
integrals used in the Sections II and III are finite. Any reasonable form
for the correlation functions can be expected to satisfy this
requirement. Although we have used only two body correlation functions
without any state dependence or backflow terms
to illustrate our results, the arguments can be easily extended to
include both of the above and also three body
correlations. Similarly, the arguments can also be extended for any
trial single particle spectrum which has a vanishing energy level
spacing in the thermodynamic limit. 

We have not addressed the problem of non diagonal matrix elements,
which are required for certain applications. One such case is
presented in the calculation of weak interaction rates, where the
relevant matrix elements are the non diagonal matrix elements of one
body operators. Work is in progress to calculate the non diagonal
matrix elements of these one body weak interaction operators including
the orthogonality corrections. 

We have also not discussed the actual forms for the correlation
functions or the trial single quasiparticle spectrum which may be
useful for calculating the free energy at finite temperature. At low
enough temperatures the zero temperature forms may provide good
approximations, but at higher temperature this is probably not
true. Methods to optimize free energy computations at finite
temperature are being developed.            
\begin{acknowledgments}      
A.M. would like to thank D. G. Ravenhall and Y. Oono for valuable
discussions. This work has been supported in part by the US NSF
via grant PHY 03-55014
\end{acknowledgments}

\newpage
\begin{table}[tbp]
\begin{tabular}{lr}
\hline
Term & Prefactor\\
\hline
$\left(\Psi_I\right|v^{eff}\left|\Psi_J \right)(\Psi_J |\Psi_I )$ & $-\frac{1}{2}$\\  
$(\Psi_I |v^{eff}|\Psi_J )(\Psi_J |\Psi_L )(\Psi_L |\Psi_I )$ &$\frac{3}{8}$\\  
$(\Psi_I |v^{eff}|\Psi_K )(\Psi_K |\Psi_L )(\Psi_L |\Psi_I )$ &$\frac{3}{8}$\\  
$(\Psi_I |v^{eff}|\Psi_J )(\Psi_J |\Psi_K )(\Psi_K |\Psi_I )$ &$\frac{3}{8}$\\  
$(\Psi_I |v^{eff}|\Psi_K )(\Psi_K |\Psi_J )(\Psi_J |\Psi_I )$ &$\frac{3}{8}$\\  
$(\Psi_I |v^{eff}|\Psi_K )(\Psi_K |\Psi_J )(\Psi_J |\Psi_K )(\Psi_K |\Psi_I )$ &$-\frac{5}{16}$\\  
$(\Psi_I |v^{eff}|\Psi_K )(\Psi_K |\Psi_J )(\Psi_J |\Psi_L )(\Psi_L |\Psi_I )$ &$-\frac{5}{16}$\\  
$(\Psi_I |v^{eff}|\Psi_K )(\Psi_K |\Psi_I )(\Psi_I |\Psi_L )(\Psi_L |\Psi_I )$ &$-\frac{5}{16}$\\  
$(\Psi_I |\Psi_L)(\Psi_L|v^{eff}|\Psi_J )(\Psi_J |\Psi_I )$ &$\frac{1}{4}$\\  
$(\Psi_I |\Psi_L )(\Psi_L |v^{eff}|\Psi_K )(\Psi_K |\Psi_I )$ &$\frac{1}{4}$\\  
$(\Psi_I |\Psi_L )(\Psi_L |v^{eff}|\Psi_J )(\Psi_J |\Psi_L )(\Psi_L |\Psi_I )$ &$-\frac{3}{16}$\\  
$(\Psi_I |\Psi_L )(\Psi_L |v^{eff}|\Psi_J )(\Psi_J |\Psi_K )(\Psi_K |\Psi_I )$ &$-\frac{3}{16}$\\  
$(\Psi_I |\Psi_L)(\Psi_L|\Psi_I )(\Psi_I |v^{eff}|\Psi_K )(\Psi_K |\Psi_I )$ &$-\frac{3}{16}$\\  
\hline 
Total&$0$\\
\hline 
\end{tabular}
\caption{The contribution of all the terms which give rise to a term like Fig.~\ref{fig3}}
\label{contrb}
\end{table} 

\newpage

\vspace{1in}

\vspace{1in}

\end{document}